\documentclass{webofc}
\usepackage[varg]{txfonts} 
\usepackage{rotating}
\usepackage{lscape}

\begin{document}
	
	\title{Characteristics of the fission fragments in the three-body model of binary fission \thanks{ 
			Y.Yu.Denisov, EPJ Web of Conferences 292, 08001 (2024); DOI: 10.1051/epjconf/202429208001}}
	
	\author{\firstname{Vitali Yu.} \lastname{Denisov}\inst{1,2,3} 
	}
	
	\institute{INFN Laboratori Nazionali di Legnaro, Legnaro(PD), Italy 
		\and
		Institute for Nuclear Research, Kiev, Ukraine 
		\and
		Faculty of Physics, Taras Shevchenko National University of Kiev, Kiev, Ukraine
	}
	
	\abstract{%
		The evaluated nuclide yields, fragment mass and charge distributions, and averaged total kinetic energy of fission fragments for the neutron-induced fission of 30 actinide nuclei are well described in the new model, which considers the fissioning scission system consisting of two heavy fragments and an $\alpha$-particle between them. The $\alpha$-particle has its origin in the neck nucleons. The yield of fission fragments in the model is linked to the number of states over the barrier of the saddle point, which is between the contacting and well-separated fission fragments. The $\alpha$-particle is fused with the nearest heavy fragment after passing the saddle point, therefore, two final fragments appear during fission. The quadrupole deformations of heavy fragments are taken into account in the model. The fragment yields depend on the heights of corresponding saddle points and the values of the equilibrium quadrupole deformation of fragments.
	}
	\maketitle
	\section{Introduction}
	\label{intro}
	The mass and charge distributions of primary fragments as well as the average total kinetic energy are important characteristics of nuclear fission. The mass dependence of the fission fragment distribution is often linked to the Gaussian \cite{vh,bgm,GEF}. The physical nature of the Gaussian behavior of the mass dependence of the fission fragment distribution is very important for an understanding of the fission process mechanism. To describe this nature Fong proposes the statistical scission-point model related to the contacting fragments with the octupole deformations \cite{fong}. The references on the modern scission-point models can be found in \cite{ds21,d22}.
	
	Just before the scission, the one-body fissioning form is similar to the two fragments connected by a neck. After neck rupture the nucleons forming the neck stay some time between the two nascent fission fragments. The neck nucleons may form one or two $\alpha$-particles, which locate between the nascent fragments. The existence of such possibilities is confirmed by the experimental observation of the ternary and quaternary fission \cite{vh,4f}. Besides this, the dynamical synthesis of $\alpha$-particle in the scission phase of nuclear fission is obtained in the time-dependent density functional model \cite{ren}. The high-density islands in the neck of the fissioning nuclei are found in the Hartree-Fock-Bogoliubov calculations \cite{han}. The formation of the $\alpha$-particles after the neck rupture may be linked to these high-density islands. The smallest radius of the neck estimated in Ref. \cite{han} is close to 2 fm. This radius is near to the radius of the $\alpha$-particle equating $\simeq 1.2 \cdot 4^{1/3} = 1.9$ fm. Moreover, the binding energy per nucleon of the $\alpha$-particle is the largest among other light nuclei of similar radii, which may be formed in the neck domain. As a result, the $\alpha$-particle leads to the lowest energy of the three-body fragment system consisting of the fragment-light nucleus-fragment \cite{ds21}. Thus, the formation of $\alpha$-particle(s) between two fragments after the neck rupture is supported both experimentally and theoretically.
	
	Recently, the new scission-point model for the description of the fission fragment characteristics has been proposed in Refs. \cite{ds21,d22}. In this model, the scission system consists of the two-, three- and four-body systems. The two-body scission systems relate to the two heavy deformed fragments. The three- and four-body scission systems consist of two heavy deformed fragments and one or two alpha-particles between them, respectively. However, the three-body scission configurations are the most important for fission, because the contribution of the four-body system in the fission process is small as a rule \cite{ds21}. This model is shortly described in Sec. 2. The results are discussed in Sec. 3 and conclusions are given in Sec. 4.
	
	\section{The model}
	
	\subsection{Two fragment systems}
	
	Let us consider the interaction of two deformed fragments with surface radii $R_i(\theta)=R_{i0}\left[1+\sum_{\ell=2}^4 \beta_{i\ell}Y_{\ell0}(\theta) \right]$. Here $R_{i0}$ is the radius of spherical fragment $i$ ($i=1,2$), $\beta_{i\ell}$ is the surface deformation parameter of fragment $i$, and $Y_{\ell0}(\theta))$ is the spherical harmonics function. The fragment's axial symmetry axis connects their mass centers, see Fig. 1, because such orientation of prolate fragments leads to the lowest barrier height of their interaction. 
	
	\begin{figure}[h]
		\centering
		\includegraphics[width=5.7cm]{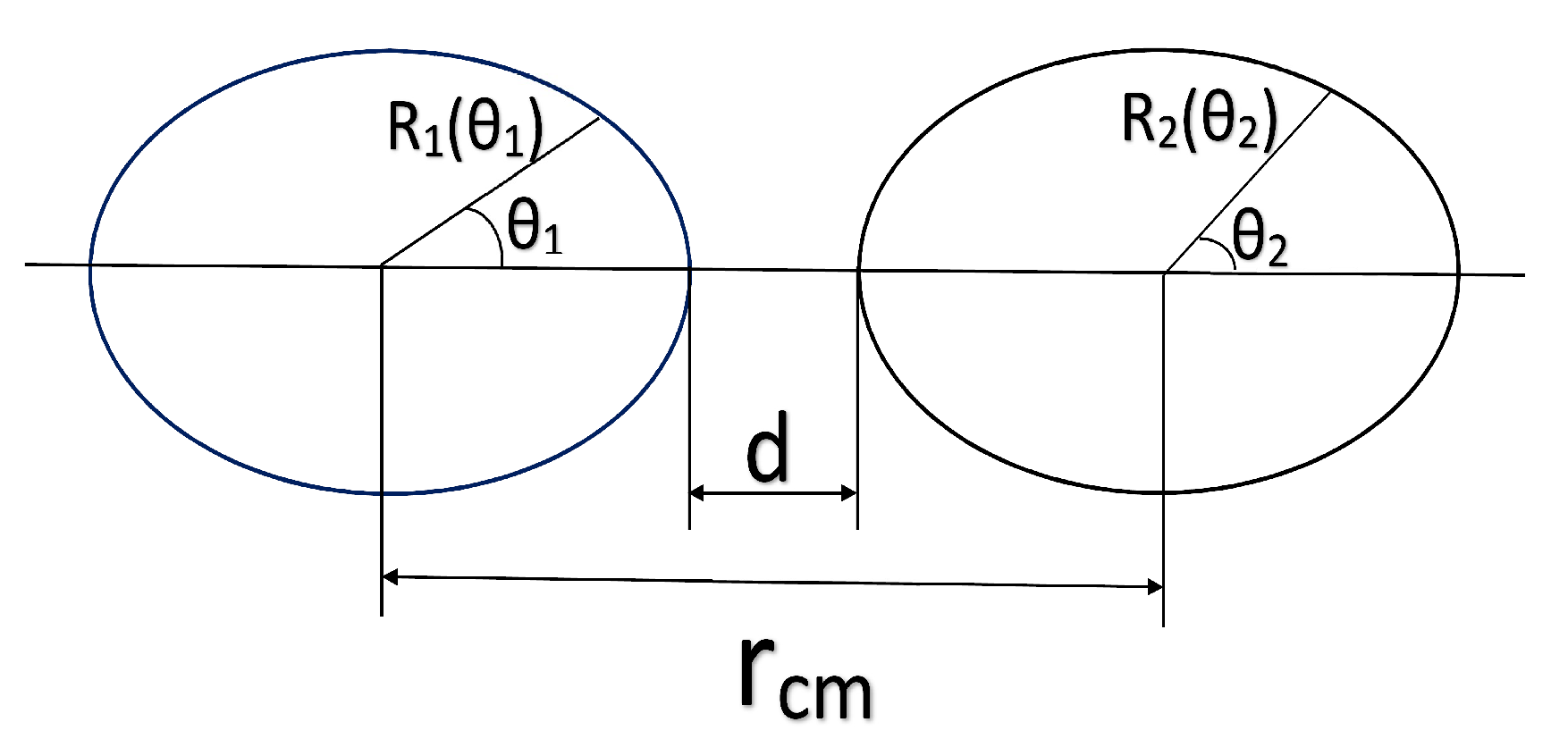} \includegraphics[width=7.0cm]{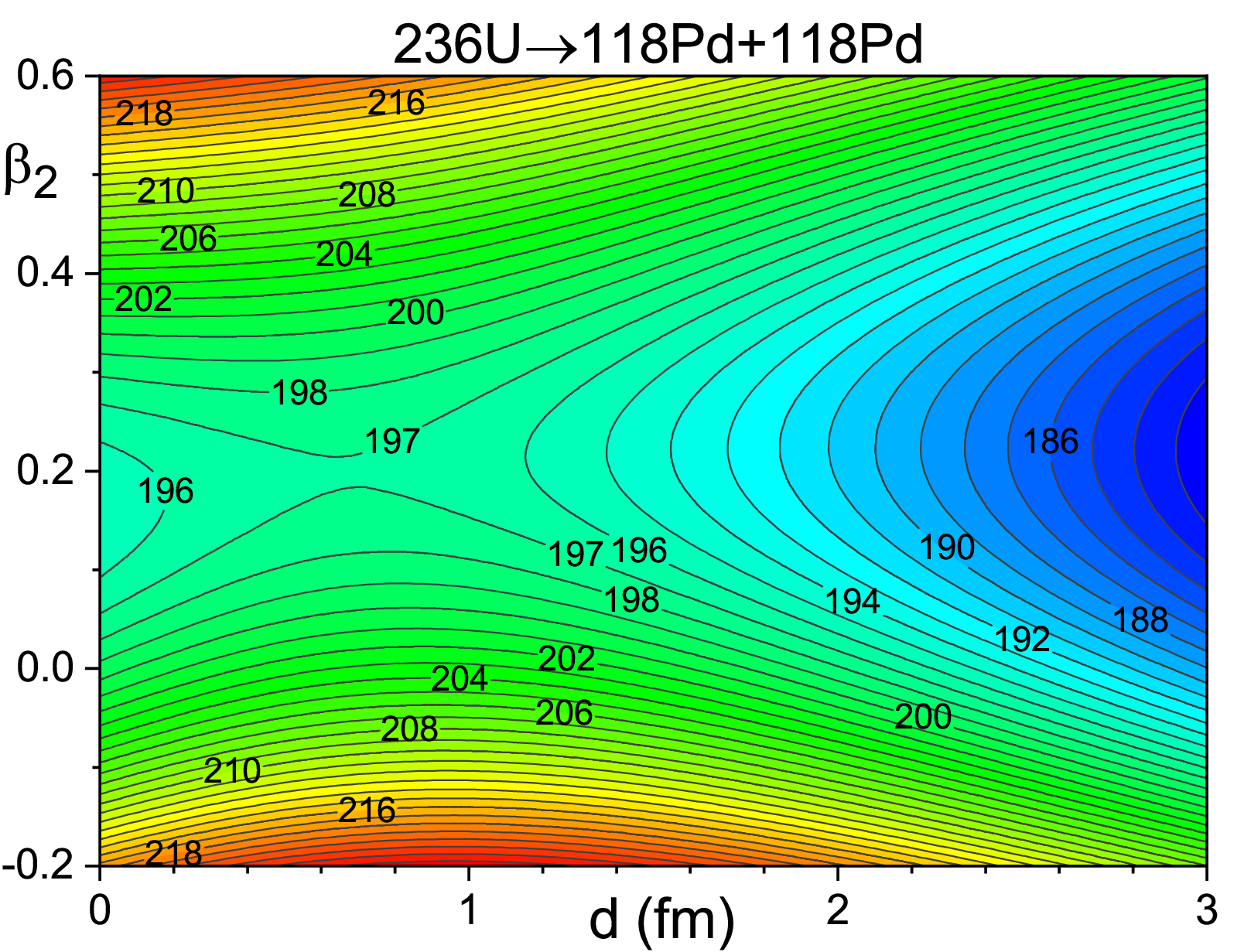}
		\caption{Left panel: The orientation of two interacting fragments. Right panel: The potential energy surface $V_{12}^{\rm tot}(d,\beta_{2})$ related to interaction of two $^{116}$Pd fragments formed at fission of $^{236}$U.}
		\label{fig-1} 
	\end{figure}
	
	The total potential energy of two deformed fission fragments after the neck rupture is
	\begin{eqnarray}
		V_{12}^{\rm tot}(r,\{ \beta_{i\ell} \}) = V^{\rm C}_{12}(r,\{\beta_{i\ell}\}) + V^{\rm n}_{12}(r,\{\beta_{i\ell}\})+E^{\rm def}_{1 \;{\rm LD}}(\{\beta_{1\ell}\}) + E^{\rm def}_{2\;{\rm LD}}(\{\beta_{2\ell}\}), 
	\end{eqnarray}
	where $V^{\rm C}_{12}(r,\{\beta_{i\ell}\})$ and $V^{\rm n}_{12}(r,\{\beta_{i\ell}\})$ are the Coulomb and nuclear interactions of the fragments, respectively, $r=r_{\rm cm}$ is the distance between the mass centers of the fragments, and $\{\beta_{i\ell}\}=\beta_{12},\beta_{22},\beta_{13},\beta_{23},\beta_{14},\beta_{24}$. $E^{\rm def}_{i\;{\rm LD}}(\{\beta_{i\ell}\})$ is the deformation energy of fragment $i$ calculated in the liquid-drop model relatively the spherical equilibrium shape. The expressions for calculation $V_{12}^{\rm tot}(r,\{\beta_{i\ell}\})$ and corresponding parameters of the potentials are given in Ref. \cite{dms}. 
	
	The potential energy surface $V_{12}^{\rm tot}(d,\beta_{2})$ related to the interaction of two $^{118}$Pd fragments dependent on the distance between closest points of the fragments $d$ and $\beta_{2}=\beta_{12}=\beta_{22}$ is presented in Fig. 1. The potential $V_{12}^{\rm tot}(d,\beta_{2})$ is found using the minimization of $V_{12}^{\rm tot}(r,\{\beta_{i\ell}\})$ on $\beta_{i3}$ and $\beta_{i4}$ for each point $(d,\beta_2)$, where $d$ is the function of $r,\beta_{12},\beta_{22},\beta_{13},\beta_{23},\beta_{14},\beta_{24}$. The height of the saddle point of the potential surface in Fig. 1 is close to 197 MeV. The Q-value of the fission reaction $^{236}$U$\rightarrow ^{118}$Pd+$^{118}$Pd obtained using the atomic mass table \cite{ame} is $Q\simeq 193$ MeV. Therefore, two contacting ($d=0$) fragments $^{118}$Pd formed in the spontaneous fission of $^{236}$U can not easily be separated, because they should overcome the barrier. 
	
	\subsection{Three fragment systems}
	
	Let's consider the interaction energy of the system consisting of two deformed fragments and an $\alpha$-particle between them \cite{ds21,d22}. The mass centers of all nuclei of this system belong to the axial-symmetry axis of the system, see Fig. 2. 
	
	\begin{figure}[h]
		\centering
		\includegraphics[width=6.6cm]{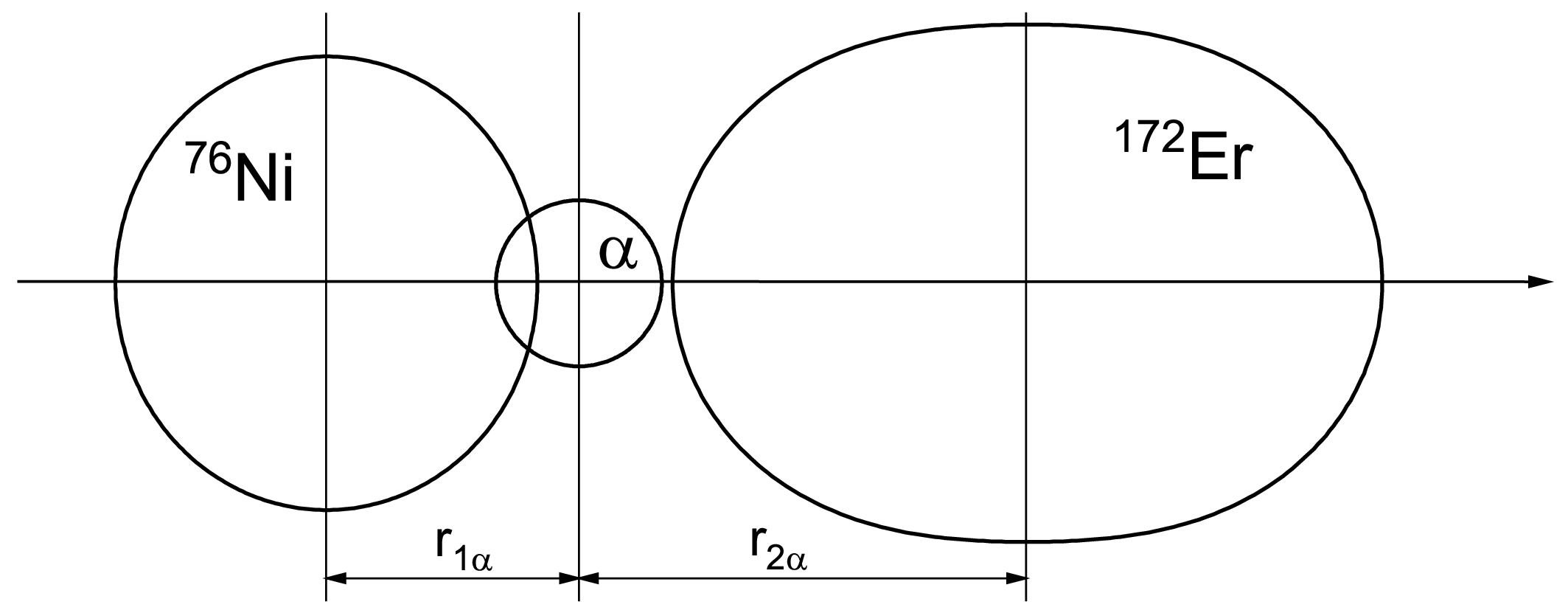} \hspace{-4mm} \includegraphics[width=6.4cm]{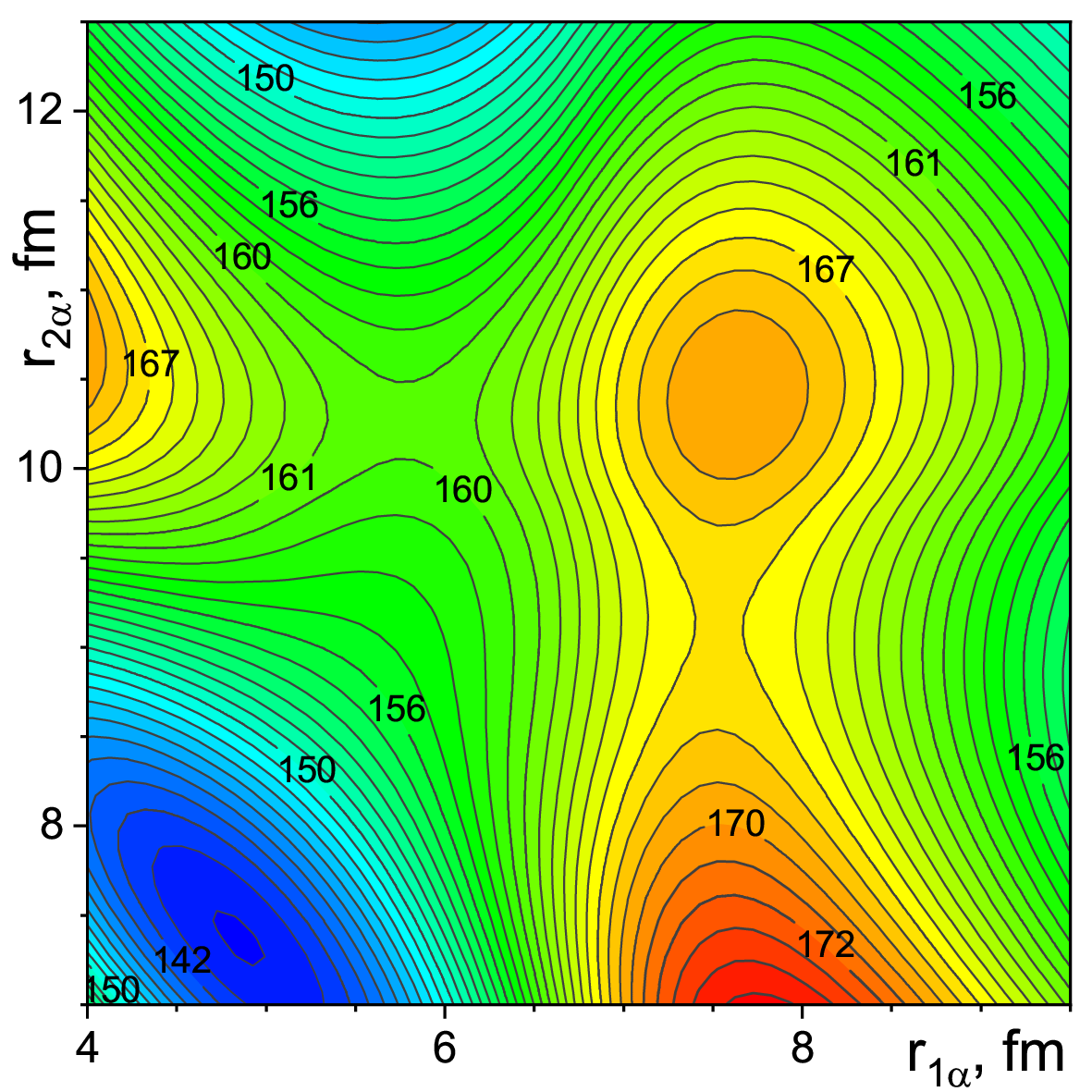}
		\caption{Left panel: The orientation of two interacting fragments and an $\alpha$-particle between them. $r_{i\alpha}$ is the distance between the mass centers of nucleus $i$ and the $\alpha$-particle. Right panel: The total interaction potential energy $V_{1\alpha2}^{\rm tot}(r_{1\alpha},r_{2\alpha},\beta_{12},\beta_{22})$ for the system ${\rm ^{76}Ni(\beta_{12}=-0.074)+\alpha+^{172}Er(\beta_{22}=0.351)}$ formed at fission of $^{252}$Cf.}
		\label{fig-2} 
	\end{figure}
	
	The total potential energy of two heavy fission fragments and an $\alpha$-particle between them is
	\begin{eqnarray} 
		V_{1\alpha2}^{\rm tot}(r_{1\alpha},r_{2\alpha},\beta_{12},\beta_{22})
		= V^{\rm tot}_{12}(r_{1\alpha}+r_{2\alpha},\beta_{12},\beta_{22}) 
		+ V^{\rm tot}_{1\alpha}(r_{1\alpha},\beta_{12}) 
		+ V^{\rm tot}_{2\alpha}(r_{2\alpha},\beta_{22}) .
	\end{eqnarray}
	Here the interaction potential of two heavy fragments $V^{\rm total}_{12}(r_{1\alpha}+r_{2\alpha},\beta_{12},\beta_{22}) $ is described in Eq. (1) at $\beta_{i3}=\beta_{i4}=0$, see, also, Refs. \cite{ds21,d22}. $V^{\rm tot}_{i\alpha}(r_{1\alpha},\beta_{12})=V^{\rm C}_{i\alpha}(r_{i\alpha},\beta_{i2})+V^{\rm n}_{i\alpha}(r_{i\alpha},\beta_{i2})$ is the total interaction potential between fragment $i$ and the $\alpha$-particle consisting of the Coulomb and nuclear parts \cite{ds21,d22}. The nuclear part of the $\alpha$-nucleus potential around the touching point is obtained using the data for the ground-state-to-ground-state $\alpha$-transition half-lives in 401 nuclei and the $\alpha$-capture cross-sections of $^{40}$Ca, $^{44}$Ca, $^{59}$Co, $^{208}$Pb, and $^{209}$Bi \cite{umadac}.

	The potential surface $V_{1\alpha2}^{\rm tot}(r_{1\alpha},r_{2\alpha},\beta_{12},\beta_{22})$ for the system ${\rm ^{76}Ni}(\beta_{12}=-0.074)$ + $\alpha$ + ${\rm ^{172}Er(\beta_{22}=0.351)}$ is presented in Fig. 2. The heights of the first $B_{1\alpha2}^{s1}(r_{\beta_1,\beta_2})$ and second $B_{1\alpha2}^{s2}(r_{\beta_1,\beta_2})$ saddle points of the potential surface $V_{1\alpha2}^{\rm tot}(r_{1\alpha},r_{2\alpha},\beta_{12},\beta_{22})$ are close to 160 and 167 MeV, respectively. The first (second) saddle for this system is taken place at distances $r_{1\alpha}\approx 5.8 (7.6)$ fm and $r_{2\alpha}\approx 10.3 (9.2)$ fm. The passing of the system through the first saddle point is related to the small changes of $r_{1\alpha}$ and large variations of $r_{2\alpha}$, which starts from a small value $r_{2\alpha}$ at the neck rupture point. The values $r_{1\alpha}$ in the position of the local minimum of the potential energy surface of the system after passing the saddle point, i.e. at a fixed value of $r_{2\alpha}> 10.4$ fm, is smaller 5.8 fm. The barrier radius of the interaction potential between the $\alpha$-particle and ${\rm ^{76}Ni}(\beta_{12}=-0.074)$ is close to 9 fm, see Fig. 3. As a result, the $\alpha$-particle is merging with ${\rm ^{76}Ni}$ after passing this saddle point because it locates in the pocket of the $\alpha$-${\rm ^{76}Ni}$ potential. Therefore, binary fission takes place generally. However, ternary fission may occur when the $\alpha$-particle overcomes the barrier of the $\alpha$-${\rm ^{76}Ni}$ potential by sub-barrier tunneling or when the fragment $^{76}$Ni accumulates high intrinsic excitation energy. Due to the sub-barrier tunneling of the $\alpha$-particle or the rare possibility of the $\alpha$-particle emission in competition to the neutron emission at high intrinsic excitation energy by the fragment $^{76}$Ni, the probability of ternary fission should be small in our model. This agrees with the experimental fact that ternary fission occurs $ \sim 10^{-3}$ times rare than binary fission \cite{vh}. 
	
	A similar analysis can be done for the second saddle point in Fig. 2 too. As a result, the $\alpha$-particle is fused with ${\rm ^{172}Er}$ after passing the second saddle point in Fig. 2.
	
		\begin{figure}[h]
		\centering
		\includegraphics[width=6.4cm]{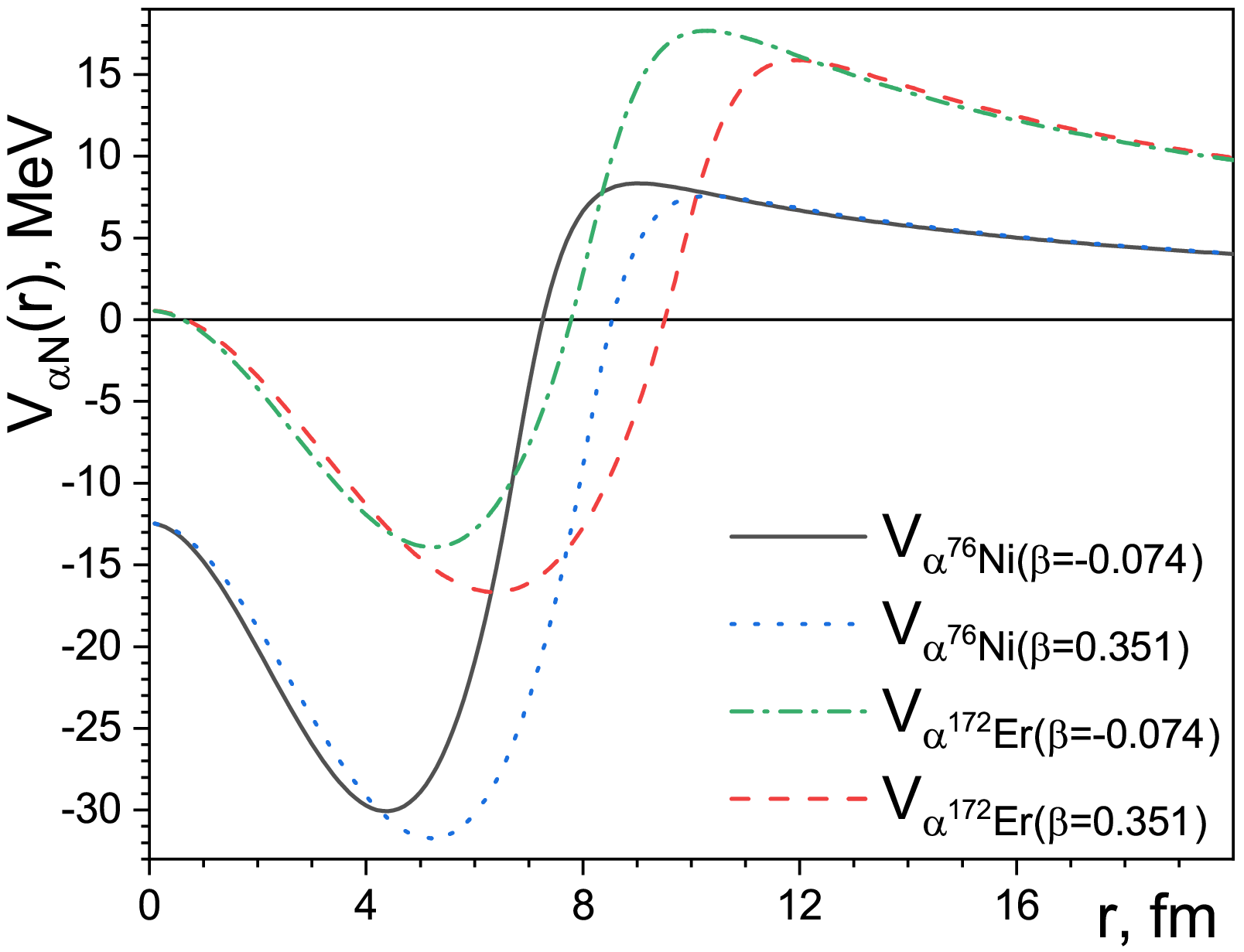}
		\caption{The radial dependencies of potentials for the systems $\alpha+^{76}$Ni and $\alpha+^{172}$Er.}
		\label{fig-3} 
		\end{figure}
	
	\subsection{The yield of the primary fission fragments}
	
	The yield of the primary fission fragments with the nucleon compositions $[A_1, Z_1]$ and $[A_2=A_{\rm cn}-A_1, Z_2=Z_{\rm cn}- Z_1]$ formed through the passing the first and second saddle points of the three-body scission configuration and normalized on 2 is 
	\begin{eqnarray}
		Y_{3}(A_1,Z_1)= 2\left[\rho_{12}^{\rm saddle \; 1}(A_1,Z_1)+\rho_{12}^{\rm saddle \; 2}(A_1,Z_1)\right]/N_{3}.
	\end{eqnarray}
	Here $A_{\rm cn}$ and $Z_{\rm cn}$ are the number of nucleons and protons in the initial fissioning nuclei,
	\begin{eqnarray}
		\rho_{12}^{\rm saddle \;1}(A_1,Z_1)=\int d\beta_{12} \int d\beta_{22} \; \rho_{A_1-4,Z_1-2}(E_1^{s1}) \; \rho_{A_2,Z_2}(E_2^{s1}),\\
		\rho_{12}^{\rm saddle \; 2}(A_1,Z_1)=\int d\beta_{12} \int d\beta_{22} \; \rho_{A_1,Z_1}(E_1^{s2}) \; \rho_{A_2-4,Z_2-2}(E_2^{s2})
	\end{eqnarray}
	are the level densities of the three-fragment system above the saddle point barriers 1 and 2 respectively,
	\begin{eqnarray}
		N_{3}= \sum_{A_1,Z_1} \left[\rho_{12}^{\rm saddle \; 1}(A_1,Z_1)+\rho_{12}^{\rm saddle \; 2}(A_1,Z_1) \right]
	\end{eqnarray}
	is the sum of all possible values $A_1$ and $Z_1$. Here $\rho_{A,Z}(U) =\frac{\sqrt{\pi} \exp{\left(2 \sqrt{a_{\rm dens}(A,Z,E) U}\right)}}{12 a_{\rm dens}(A,Z,E)^{1/4} U^{5/4}}$ is the back-shifted Fermi gas level density \cite{ripl3}, where $U=E-\Delta$ and $a_{\rm dens}(A,Z,E) = a_{\rm inf} \left[1 + \delta_{\rm shell} (1 -\exp{(-\gamma U)})/U \right]$. The parameters of the level density $a_{\rm inf}$, $\delta_{\rm shell}$, $\gamma$, and $\Delta$ are taken from Ref. \cite{ripl3} without any changes. $E_1^{s1}$, $E_2^{s1}$, $E_1^{s2}$, and $E_2^{s2}$ are the thermal excitation energies of the nascent fragments. $\alpha$-particle is the inert particle. The temperatures of the fragments are the same ($T_1^{s_k} = T_2^{s_k}$) in each saddle point, therefore, there are two systems of equations for finding $E_1^{s1}$, $E_2^{s1}$, and $E_1^{s2}$, $E_2^{s2}$, respectively, 
	\begin{eqnarray}
		\left\{\begin{split}
			\frac{E_1^{s1}-\Delta_1^{s1}}{a_{\rm dens}(A_1,Z_1,E_1^{s1})} \equiv (T_1^{s1})^2 = (T_2^{s1})^2 \equiv \frac{E_2^{s1}-\Delta_2^{s1}}{a_{\rm dens}(A_2,Z_2,E_2^{s1})} , \\
			E_1^{s1}+E_2^{s1}=Q_3^{s1}+E_{\rm cn}-B_{1\alpha2}^{s1}(r_{\beta_1,\beta_2}), \\
		\end{split}\right. \\
		\left\{\begin{split}
			\frac{E_1^{s2}-\Delta_1^{s2}}{a_{\rm dens}(A_1,Z_1,E_1^{s2})} \equiv (T_1^{s2})^2 = (T_2^{s2})^2 \equiv \frac{E_2^{s2}-\Delta_2^{s2}}{a_{\rm dens}(A_2,Z_2,E_2^{s2})}, \\
			E_1^{s2}+E_2^{s2}=Q_3^{s2}+E_{\rm cn}-B_{1\alpha2}^{s2}(r_{\beta_1,\beta_2}).\\
		\end{split}\right.
	\end{eqnarray} 
	Here $B_{1\alpha2}^{s1}(r_{\beta_1,\beta_2})$ and $B_{1\alpha2}^{s2}(r_{\beta_1,\beta_2})$ are the height of the saddle point $i$ of the total potential energy $V_{1\alpha2}^{\rm tot}(r_{1\alpha},r_{2\alpha},\beta_{12},\beta_{22})$ (see Fig. 2), $E_{\rm cn}$ is the initial excitation energy of the fissioning nucleus, and $Q_3^{si}$ is the Q-value of fission of initial nucleus into the three-fragment system calculated using \cite{ame}.
	
	\subsection{ Steps for the calculation of the yields of fragments in the model}
	The sequential steps for finding the yields of fragments in the model are:\\
	1. Calculation of Q-value of fission of initial nucleus into fragment-$\alpha$-fragment system with numbers of the nucleons and protons in the first fragment $A_1, Z_1$.\\
	2. Calculation of the saddle point heights $B_{1\alpha2}^{s1}(r_{\beta_1,\beta_2})$ and $B_{1\alpha2}^{s2}(r_{\beta_1,\beta_2})$ for the selected fragment partitions at given values of the fragment deformations $\beta_i$ using Eq. (2). \\
	3. Calculation of the excitation energy of each fragment $E_k^{s i}$ by equating the temperatures of fragments applying Eqs. (7) or (8) in dependence on the chosen saddle point. \\
	4. Calculation of the product of the level densities of the system $\rho_{A_1, Z_1}(E_1^{s i}) \rho_{A_2, Z_2}(E_2^{s i})$ presented in the integrals (4)-(5) using the fragment excitation energies $E_k^{s i}$. \\ 
	5. Calculation of $\rho_{12}^{\rm saddle \; i}(A_1, Z_1)$ by taking the integrals (4) and (5) on $\beta_{12}$ and $\beta_{22}$. \\
	6. Repeat steps 1-5 for all possible fission fragment variants $A_1, Z_1$. In practice, the set of values of $A_1, Z_1$ may be limited by the set of all experimentally selected yields of nuclei that can occur in 30 considered induced fission reactions. \\ 
	7. Calculation of the norm $N_{3}$ using Eq. (6). \\ 
	8. Calculation of the fragment yield $Y(A, Z)$ with the help of Eq. (3). The mass and charge distributions of fission fragments are found by applying expressions $Y(A)=\Sigma_Z Y(A, Z)$ and $Y(Z)=\Sigma_A Y(A, Z)$, respectively. 
	
	So, the calculations of the yields $Y(A, Z)$, $Y(A)$, and $Y(Z)$ are straightforward. 
	
	\section{Discussion}
	
	The deformation energy of the fragment evaluated relative to the ground state with the quadrupole deformation $\beta_0$ is approximated in the model as
	\begin{eqnarray}
		E_{\rm def}(\beta) = (1/2)(C_{\rm ld}+C_{\rm sc}) \; (\beta-\beta_0)^2,
	\end{eqnarray}
	where $C_{\rm ld}$ is the liquid-drop stiffness, $C_{\rm sc} \approx - 0.05 \cdot C_{\rm ld} \cdot \delta_{\rm shell}$ is the shell-correction stiffness, and $\delta_{\rm shell}$ is the phenomenological value of the shell-correction \cite{ds21}. The height of the saddle point of the total potential (2) depends on the deformation energy and, therefore, it depends on $\beta_0$. As a result, the yield of the fission fragments is sensitive to the values of $\beta_0$. The equilibrium quadrupole deformation of each fragment is taken into account in the model. 
	
	There are 26461 evaluated fragment yields $Y_{\rm exp}(A,Z)$ of fission fragments formed in neutron-induced fission reactions by $^{227,229,232}$Th, $^{231}$Pa, $^{237,238}$Np, $^{232,233,234,235,236,237,238}$U, $^{238,239,240,241,242}$Pu, $^{241,243}$Am, $^{242,243,244,245,246,248}$Cm, $^{249,251}$Cf, $^{254}$Es, and $^{255}$Fm in JENDL \cite{jendl}. The values of $Y_{\rm exp}(A,Z)$ belong to the range $ 10^{-21} \lesssim Y_{\rm exp}(A,Z)\lesssim 10^{-1}$. 9020 relatively well-defined evaluated values of $Y_{\rm exp.}(A,Z)$ are selected by applying the rules: $Y_{\rm exp.}(A,Z) \geq 10^{-8}$, $Y_{\rm exp.}(A,Z) \geq 10^{-3} Y_{\rm exp.}(A) = 10^{-3} \sum_Z Y_{\rm exp.}(A,Z)$, and $Y_{\rm exp.}(A_2=A_{\rm cn}-A,Z_2=Z_{\rm cn}- Z) \geq 10^{-8}$. 
	All selected fragments are taken into account in the calculations of the fragment yields.
	
	The comparison of the mass $Y(A)$ and charge $Y(Z)$ distributions of fission fragments for the reactions $n_{\rm th.}+^{235}$U$\rightarrow^{236}$U$\rightarrow f$ and $n_{\rm 0.5 MeV}+^{238}$Pu$\rightarrow^{239}$Pu$\rightarrow f$ 
	calculated in the model with the evaluated data \cite{jendl} is presented in Fig. 4. Results are shown on both linear and logarithmic scales to provide complementary information. Figures for the other 28 nuclei similar to the ones presented in Fig. 4 are given in Ref. \cite{d22}. The comparison of isotopes yields $Y(A, Z)$ calculated in the model for the reaction $n_{\rm 0.5 MeV}+^{238}$Pu$\rightarrow^{239}$Pu$\rightarrow f$ with the evaluated data is presented in Fig. 5. The model results agree well with the evaluated data. 
	
	Due to the energy conservation, the kinetic energy of the fragments at the infinite distance can link to the interaction potential energy of fragments and $\alpha$-particle at the corresponding saddle point \cite{ds21,d22}. The kinetic energies of the fragments on the infinite distance in the mass centers formed by passing the saddle point $i$ ($i=1,2$) are
	$ E_{\rm kin}^{si}= V^{{\rm tot}}_{12}(r_{1\alpha}^{si}+r_{2\alpha}^{s1},\beta_1^{si},\beta_2^{si}) + V_{j\alpha}^{\rm tot}(r_{j\alpha}^{si},\beta_j^{si})$.
	Here all interaction potentials are evaluated at the corresponding saddle points, $j=1,2$, and $i \neq j$. The potentials presented in this equation 
	are defined in Eqs. (1) and (2). The fusion of the $\alpha$-particle with the nearest nuclei is taken into account. The average total kinetic energy ($\overline{\rm TKE}$) is
	\begin{eqnarray}
		\overline{\rm TKE} = \frac{1}{N_3} \sum_{A_1,Z_1} \int d\beta_1 \int d\beta_2 \left[ \rho_{A_1-4}(E_1^{s1}) \rho_{A_2}(E_2^{s1}) E_{\rm kin}^{s1} + \rho_{A_1}(E_1^{s2}) \rho_{A_2-4}(E_2^{s2}) E_{\rm kin}^{s2} \right] .
	\end{eqnarray}
	\begin{figure}[h]
		\centering
		\includegraphics[width=3.16cm]{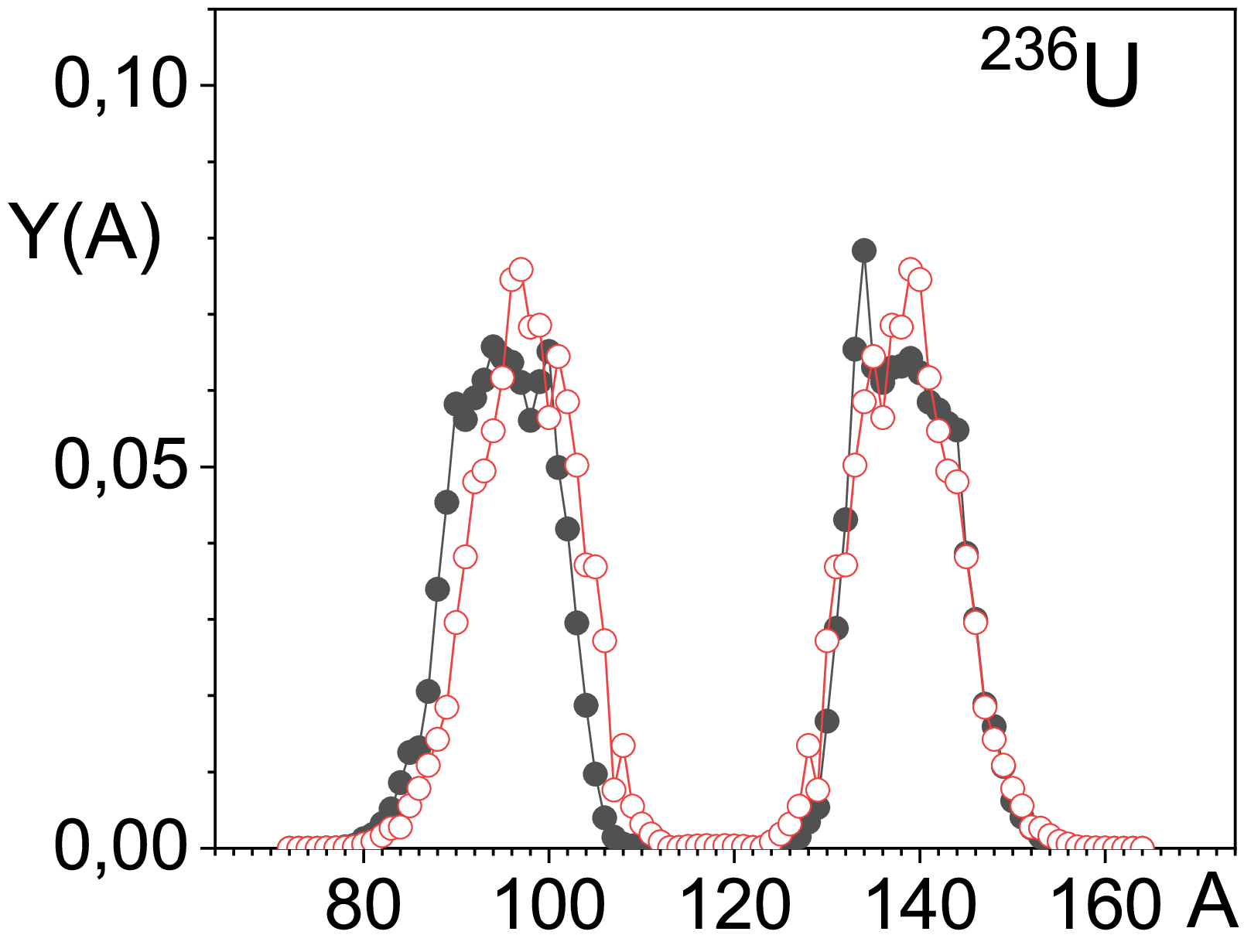}
		\includegraphics[width=3.16cm]{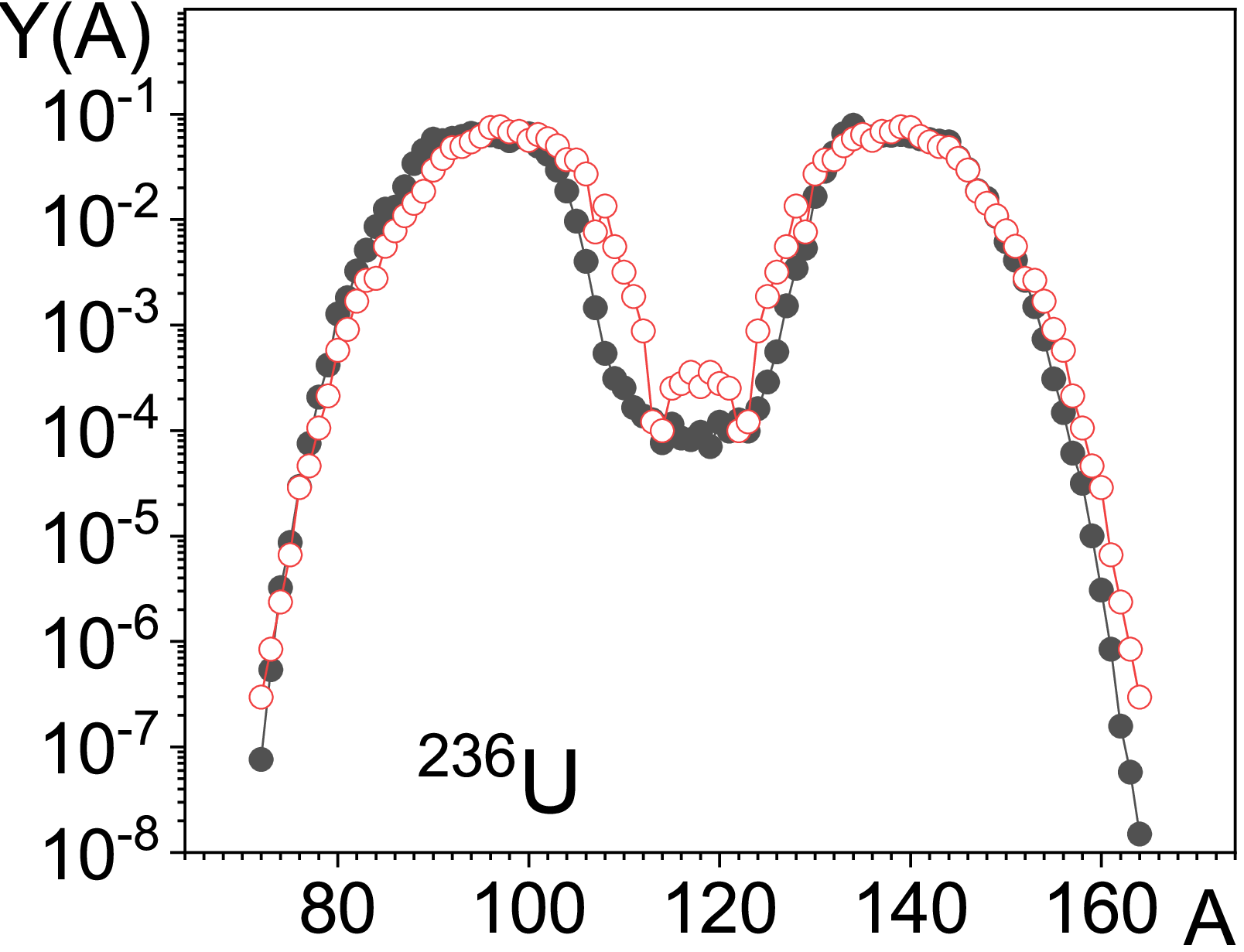} 
		\includegraphics[width=3.16cm]{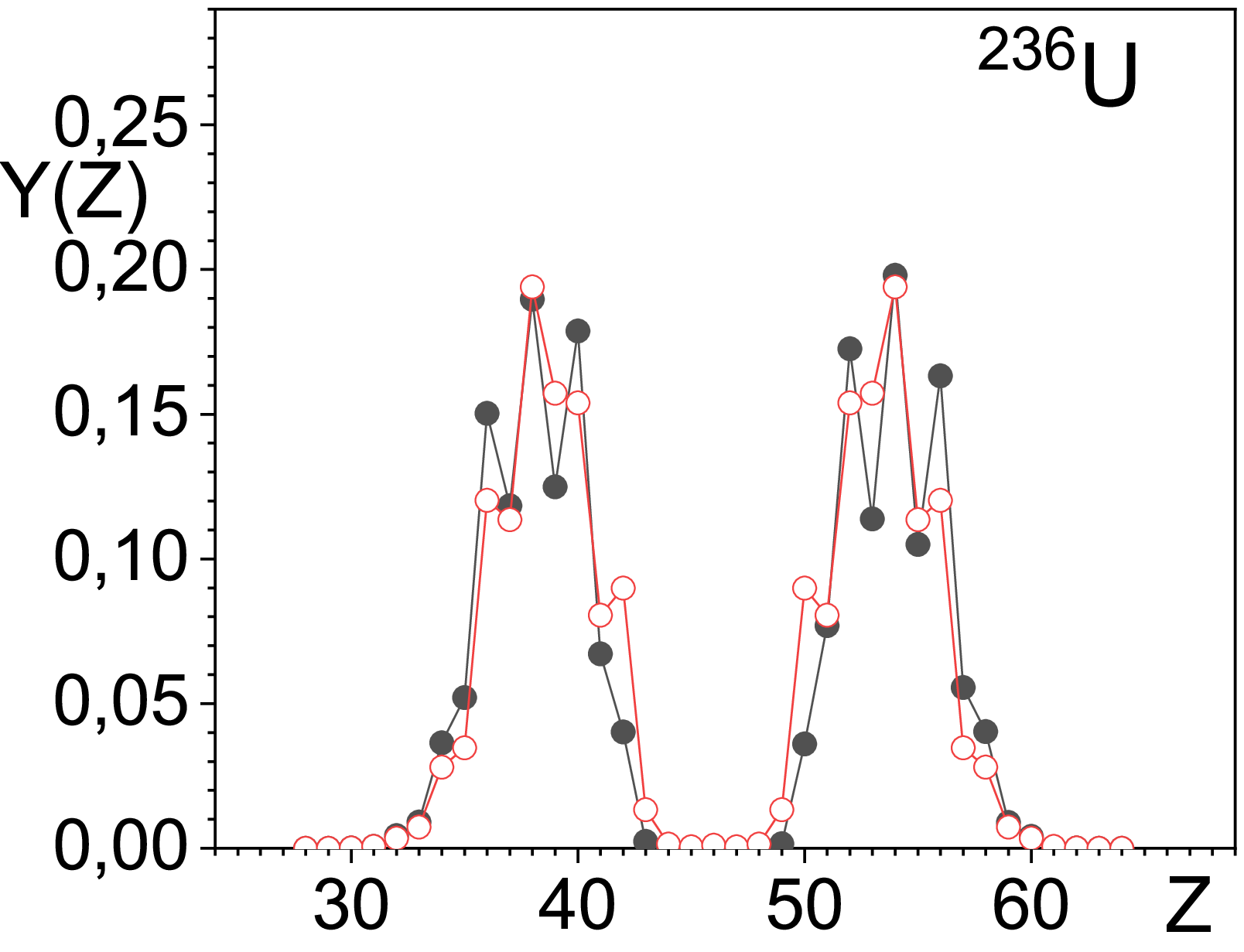}
		\includegraphics[width=3.16cm]{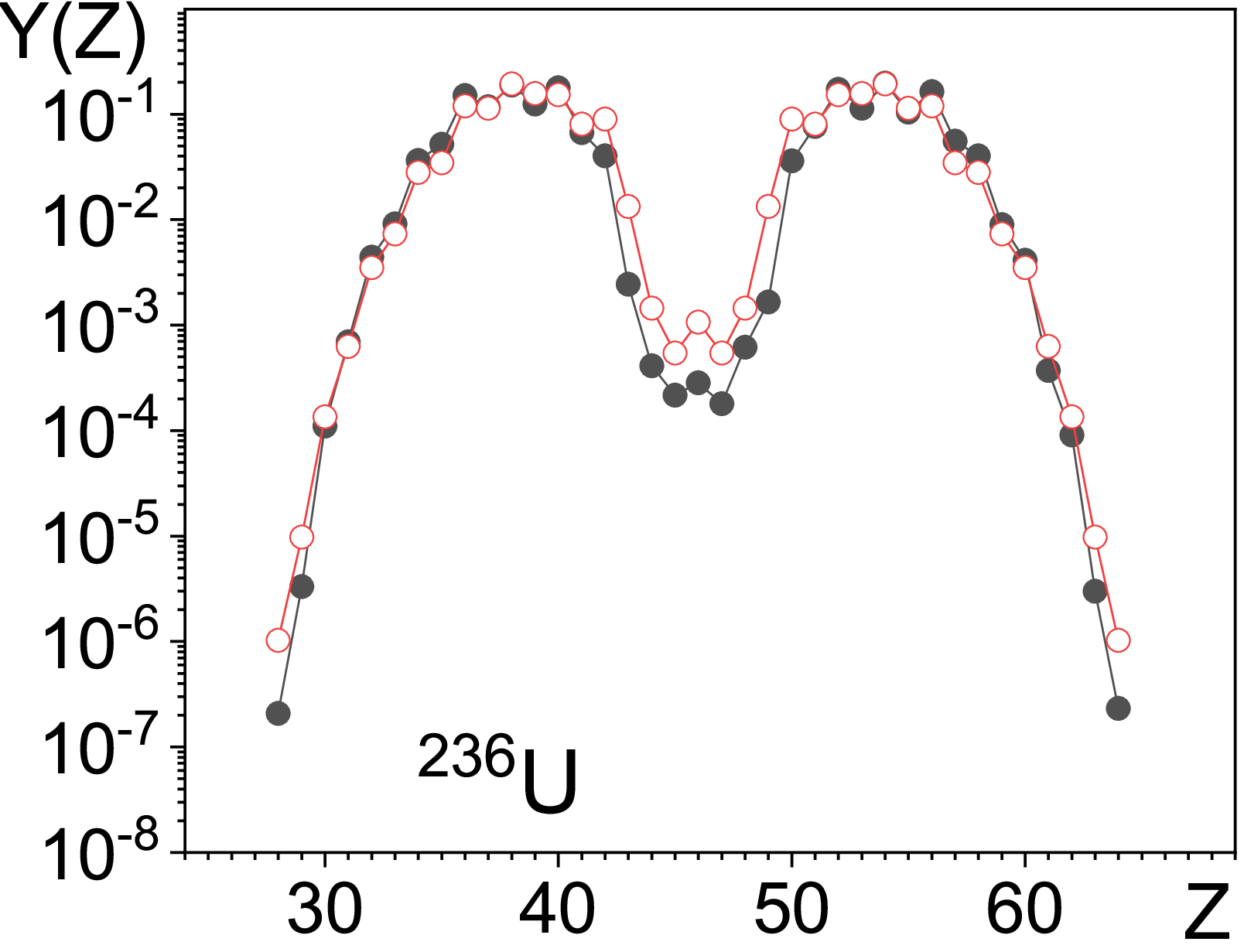}
		\includegraphics[width=3.16cm]{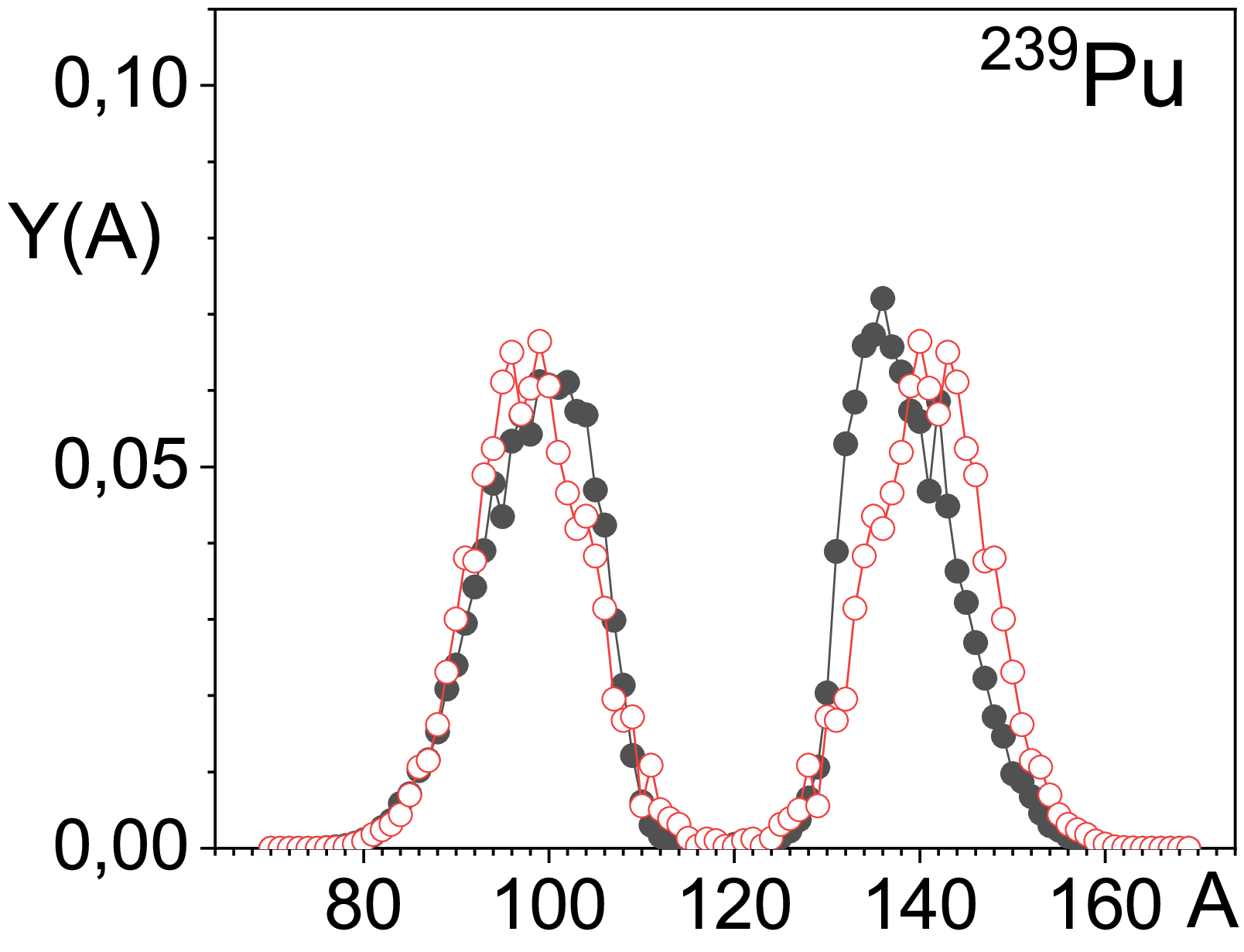} \includegraphics[width=3.16cm]{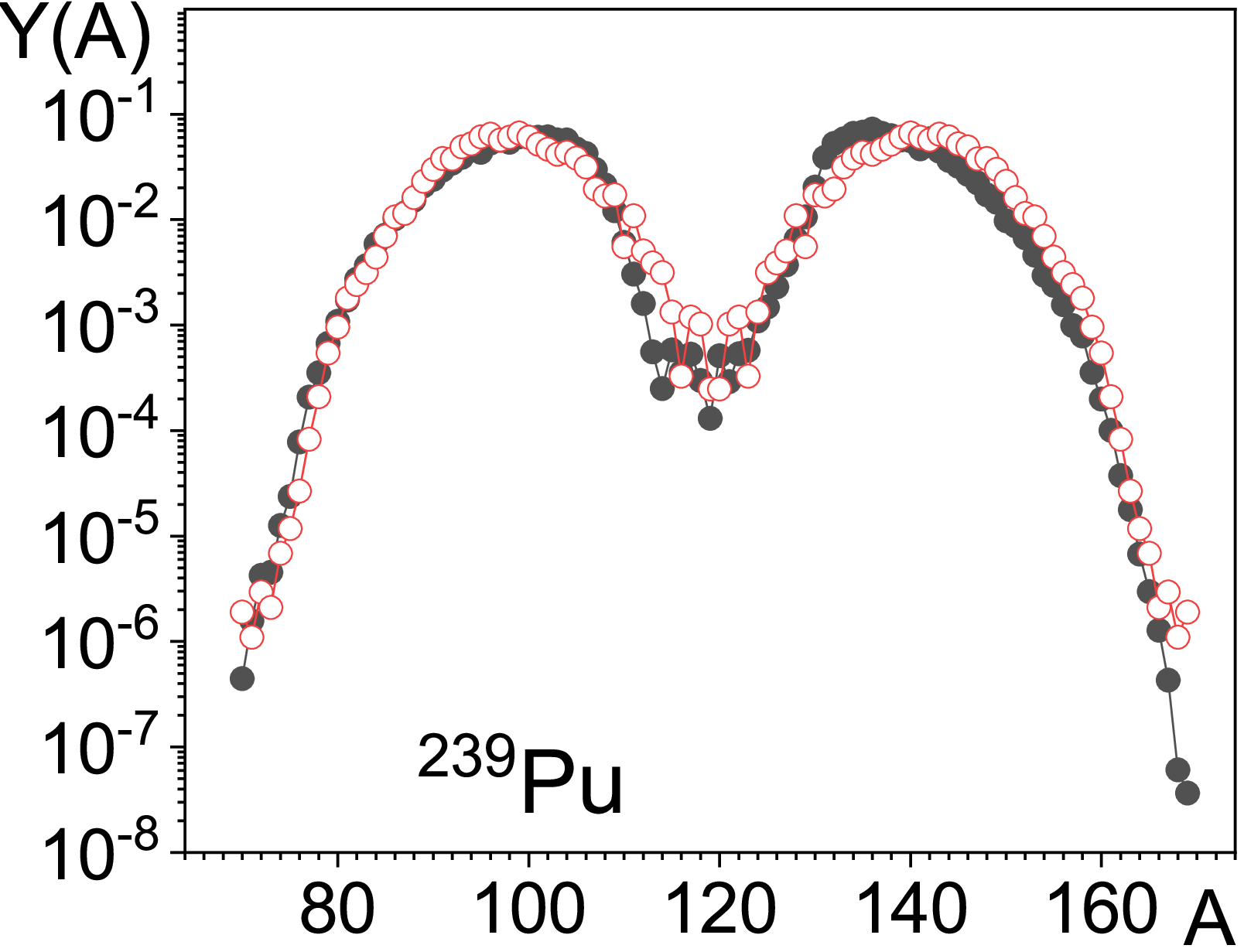} 
		\includegraphics[width=3.16cm]{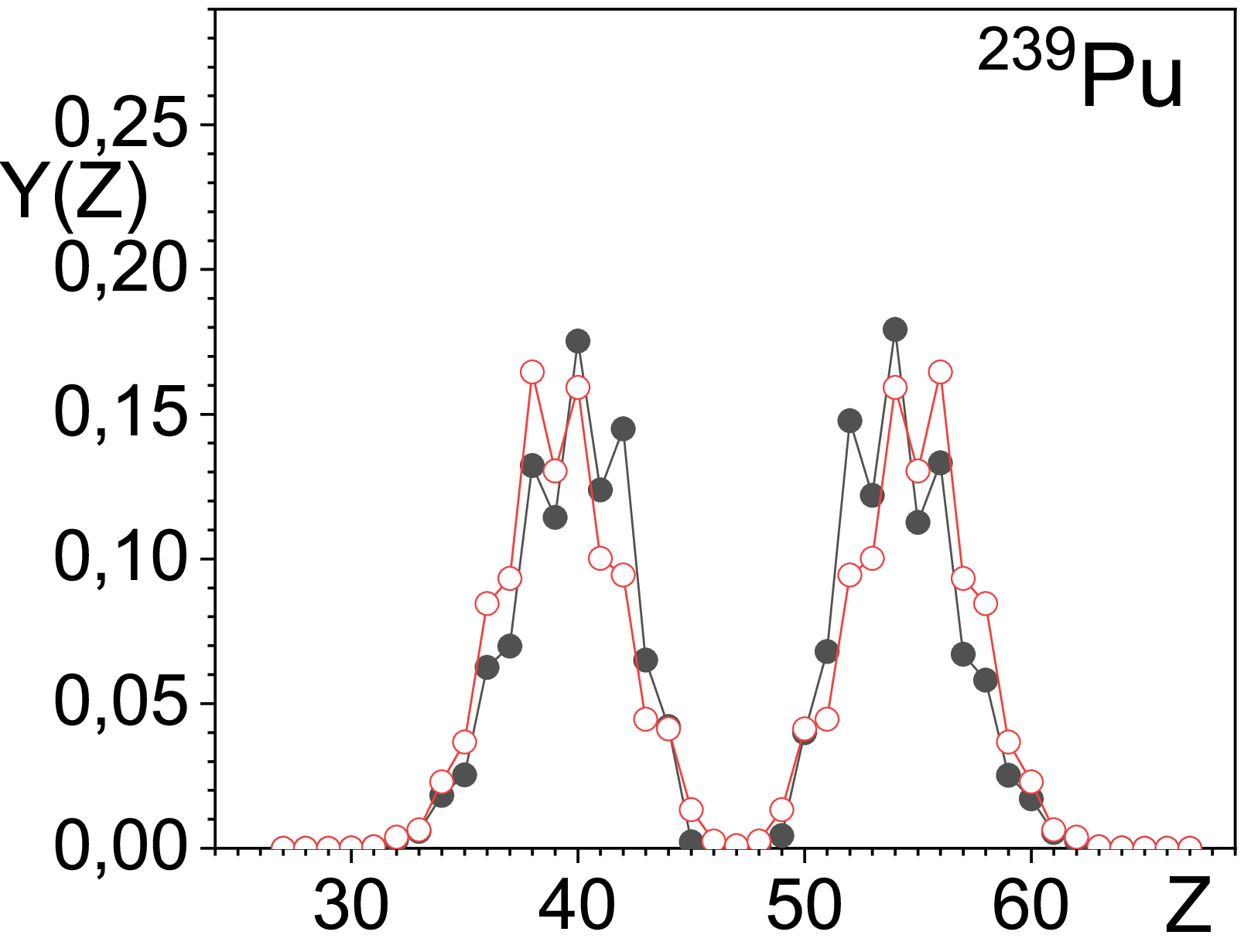} 
		\includegraphics[width=3.16cm]{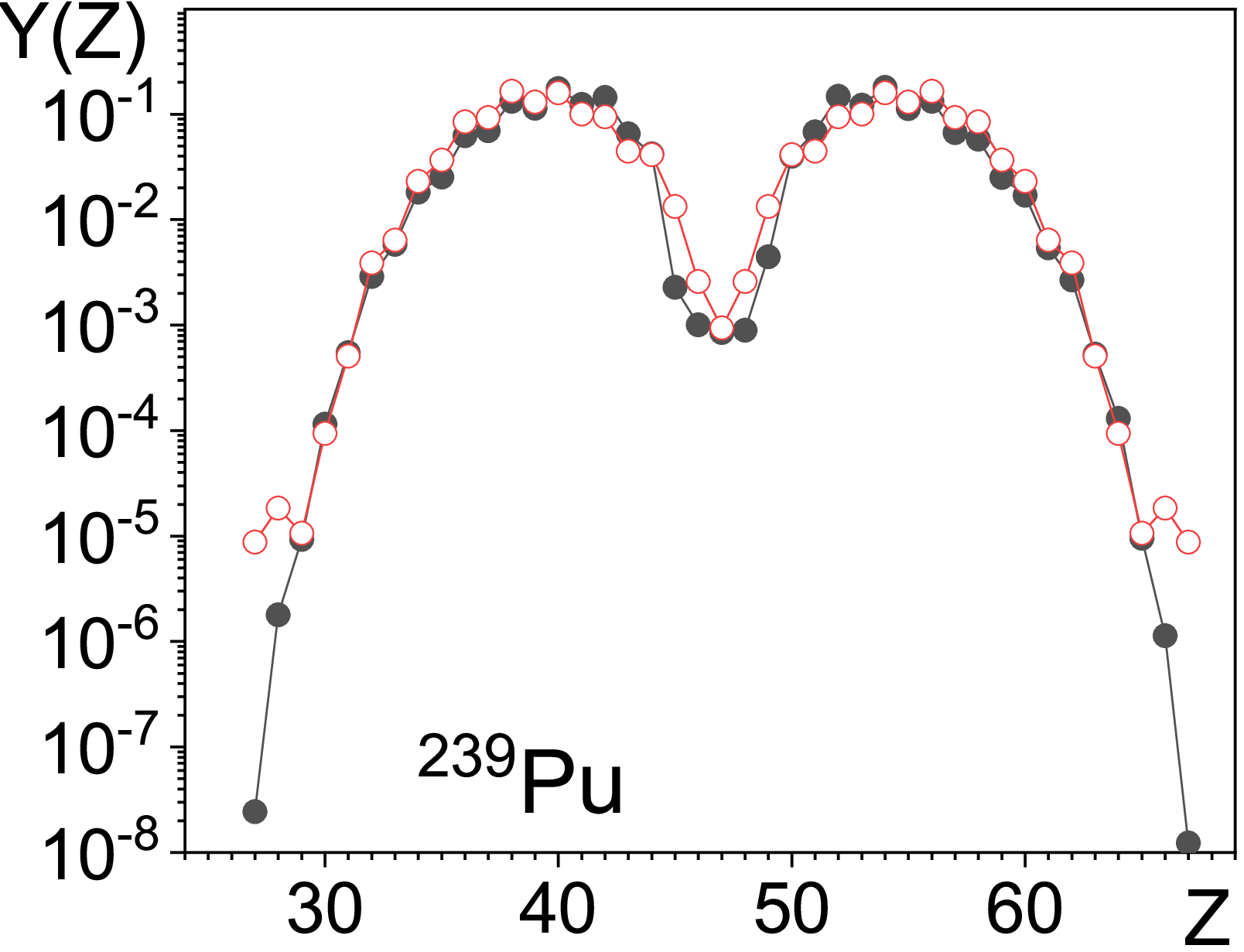}
		\caption{Comparison of the mass $Y(A)$ and charge $Y(Z)$ distributions of fission fragments for the reactions $n_{\rm th.}+^{235}$U$\rightarrow^{236}$U$\rightarrow f$ (upper row) and $n_{\rm 0.5 MeV}+^{238}$Pu$\rightarrow^{239}$Pu$\rightarrow f$ 
			(bottom row)
			in linear and logarithmic scales calculated in the model (open dots) with the evaluated data (filled dots) from JENDL \cite{jendl}.}
		\label{fig-5} 
	\end{figure}
	
	\begin{landscape}
		\begin{figure}
			\includegraphics[width=21.0cm]{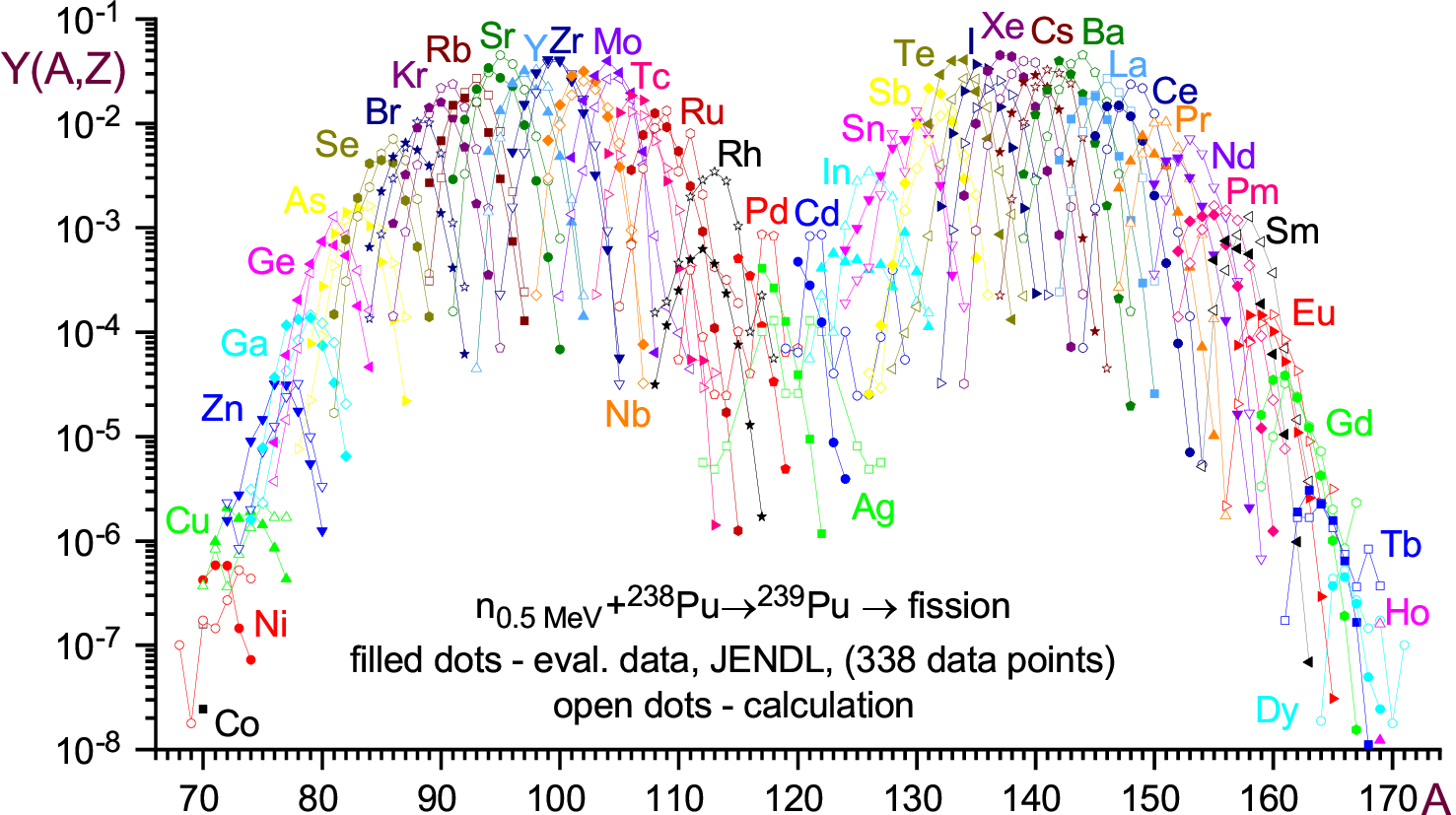}
			\caption{Comparison of the isotope yields $Y(A, Z)$ for the reaction $n_{\rm 0.5 MeV}+^{238}$PuU$\rightarrow^{239}$Pu$\rightarrow f$ calculated in the model (open dots) with the evaluated data from JENDL (filled  dots) \cite{jendl}.}
			\label{fig-4} 
		\end{figure} 
	\end{landscape}
	
	The values of $\overline{\rm TKE}$ calculated in the model without the introduction of additional parameters for the neutron-induced fission of the actinides are given in Table 1. 
	The values of $\overline{\rm TKE}$ evaluated in the model agree well with the available experimental data, see Table 1 and Ref. \cite{d22} for details.
	
	\begin{table}
		\centering
		\caption{The experimental and theoretical values of $\overline{\rm TKE}$ of primary fragments (in MeV) formed in the neutron-induced fission reactions.}
		\label{tab-1} 
		\begin{tabular}{lll|lll}
			\hline
			Reaction & $\overline{\rm TKE}_{\rm theory}$ & $\overline{\rm TKE}_{\rm exp.}$ & Reaction & $\overline{\rm TKE}_{\rm theory}$ & $\overline{\rm TKE}_{\rm exp.}$ \\
			\hline
			$n_{\rm th.}+^{229}$Th & 167.9 & 163.6$\pm$0.5 &
			$n_{\rm 0.5 MeV}+^{232}$Th & 167.8 & 162 \\ 
			$n_{\rm 0.5 MeV}+^{231}$Pa & 170.3 & 166.6$\pm$0.5 &
			$n_{\rm th.}+^{232}$U & 173.2 & 170.7$\pm$0.5 \\ 
			$n_{\rm th.}+^{233}$U & 173.0 & 171.5$\pm$0.2 &
			$n_{\rm 0.5 MeV}+^{234}$U & 173.0 & 170.58$\pm$0.05 \\ 
			$n_{\rm th.}+^{235}$U & 172.8 & 170.6$\pm$0.6 &
			$n_{\rm 0.5 MeV}+^{236}$U & 172.8 & 170.5 \\ 
			$n_{\rm 0.5 MeV}+^{238}$U & 172.7 & 170.2$\pm$1.4 &
			$n_{\rm th.}+^{237}$Np & 175.2 & 174.7$\pm$0.6 \\ 
			$n_{\rm 0.5 MeV}+^{238}$Pu & 177.8 & 179.7$\pm$0.5 & 
			$n_{\rm th.}+^{239}$Pu & 177.7 & 178.8$\pm$0.5 \\ 
			$n_{\rm th.}+^{240}$Pu & 177.8 & 178.2$\pm$0.5 & 
			$n_{\rm th.}+^{241}$Pu & 177.5 & 179.00$\pm$0.06 \\ 
			$n_{\rm th.}+^{242}$Pu & 177.6 & 179.32$\pm$1.8 &
			$n_{\rm th.}+^{241}$Am & 180.4 & 181.6$\pm$0.4 \\ 
			$n_{\rm th.}+^{245}$Cm & 182.8 & 188.5$\pm$3.0 &
			$n_{\rm th.}+^{249}$Cf & 188.3 & 187.3$\pm$1.5 \\
			\hline
		\end{tabular}
	\end{table}
	
	\section{Conclusions}
	
	The model for the description of the binary fission based on the three-body scission configuration is proposed. The three-body scission configuration consists of two heavy fragments and an $\alpha$-particle located between them. Such a description of the scission configuration makes it possible to describe the processes occurring in the neck during fission. There are the saddle points between touching and well-separated heavy deformed fragments of the three-body system. The yield of the fragment $Y(A, Z)$ is proportional to the number of levels at the saddle points. After passing through the saddle point the $\alpha$-particle is fused with the nearest nucleus. The values of the ground-state deformation parameters of heavy fragments are defined in the description of the nuclide yields. The yields of fragments $Y(A, Z)$, the mass $Y(A)$ and charge $Y(Z)$ fission fragment distributions, as well as, the average total kinetic energy $\overline{\rm TKE}$ of the fragments are well described in the model simultaneously.
	

\end{document}